\documentclass[aps,prl,twocolumn,groupedaddress]{revtex4}
\begin{document}


\title{Analyzing Correlations for spin-$\frac{1}{2}$ particles and Singlet pairs} 


\author{Charles Tresser}
\email[]{charlestresser@yahoo.com}
\affiliation{IBM, P.O.  Box 218, Yorktown Heights, NY 10598, U.S.A.}


\date{\today}

\begin{abstract}
We review the computation of correlations of successive projections of the spin onto axes for spin-$\frac{1}{2}$ particles and EPRB pairs (in the Singlet state). We assume forms of Realism (at least as general as the Predictive Hidden Variables in the classical Bell's Theory), to review one particle correlations before we further assume Locality and analyze EPRB pairs. To our surprise, we find that the Abstract Bell Inequalities with three or four axes cannot have physical meaning \textit{i.e.,} two spin projections cannot belong to one particle or some correlations that one would need cannot be evaluated.  A recent version of Bell's Theory without Locality turns out to be unaffected by our results that refocus the open issues on ``realism" rather than ``local realism", inviting us to leave locality alone so to speak.        
\end{abstract}
%
\maketitle

%
%
%
%

\noindent
\textbf{1)}  \textbf{Introduction.}  We consider neutral spin-$\frac{1}{2}$ particles (\textit{e.g.,} neutrons) and \emph{EPRB pairs} \cite{EPR}, \cite{Bohm}, \textit{i.e.,} \emph{entangled pairs} (soon to be defined) of such particles, pairs whose spin part of the wave function is the \emph{Singlet state}:
\begin{equation}\label{Singlet}
\Psi_{\rm Spin}(x_p,x_{p'})=\frac{1}{\sqrt{2}}(| +\rangle _p\otimes| -\rangle_{p'}-| -\rangle_p\otimes| +\rangle_{p'})\,,
\end{equation}
where $x_p$ and $x_{p'}$ are respectively the locations attributed to the particles $p$ and $p'$ of the pair $(p,\,p')$ when one evaluates $\Psi_{\rm Spin}$. 
We measure spin projections in $\frac{\hbar}{2}$ units so that for spin-$\frac{1}{2}$ particles measurements outputs are in the set $\{-1,\,1\}$. If a spin projection that is measured on either $p$ or $p'$ of some EPRB pair$(p,\,p')$  or on an individual particle is not one of $1$ or $-1$, then the pair $(p,\,p')$ or the individual particle is not taken into account.  The spin state (\ref{Singlet}) is an example of \emph{entanglement} (hence the qualification of \emph{``entangled"} that is stated above), \textit{i.e.,} the sum of tensor products in (\ref{Singlet}) cannot be rewritten as one tensor product of one-particle states. 

\smallskip
More precisely we consider $i$-indexed sequences of spin-$\frac{1}{2}$ particles  and then of EPRB pairs.  The particles of the pair indexed by $i$ we call $p_i$ and $p'_i$ but we usually avoid using the index (as we did already in (\ref{Singlet})) and rather work with a \emph{typical element} of the sequence.  So for instance if for all $i=1,2,\dots$, the spins of $p_i$ are projected along a constant \emph{axis} $d$ (an oriented straight line) at times $t_1<t_2<\dots$, we can speak of \emph{the spin projection sequence $s(p,d)=(s(p_1,d,t_1),\,s(p_2,d,t_2),\,\dots $ of $p$ along $d$}, thus mixing a constant axis $d$ with a sequence of distinct particles that we nevertheless represent by the constant symbol $p$.  We find it convenient to use $t_0$ and $t'_0$ as generic time values, with, \textit{e.g.,} $t_0$ the generic symbol for $t_k$. 

\smallskip
EPRB pairs have become important because of the impact in and much beyond the foundations of Quantum Mechanics (or \emph{QM}) both of Bell's Theory (see, \textit{e.g.,} \cite{Bell}, \cite{CHSH69}, \cite{Wigner70}, \cite{Stapp1971}, as well as \cite{Shimony}, \cite{Le Bellac} and references therein)  and of experiments about it (see, \textit{e.g.,} \cite{AspectEtAl1982},\cite{Shimony} and references therein).  QM is not adapted \textbf{by itself} to develop Bell's Theory since given any EPRB pair, QM only lets us consider the correlation of projections of the spins of the two particles each along one arbitrary axis. We follow the frequent usage (but see, \textit{e.g.,} \cite{Wigner70}) that all axes, say $a$ for $p$ and $a'$ for $p'$ (\textit{i.e.,} one axis per particle because of the Uncertainty Principle), be in a plane  orthogonal (or about so because of unavoidable lack of precision) to the Classical trajectories . In particular, as a consequence of (\ref{Singlet}), QM tells us that, with $u(d)$ standing for the unit vector along axis $d$, the \emph{correlation} of $s(p,a)$ and $s(p',a')$ is (with probability 1) the limit $-u(a)\cdot u(a')$ of a convergent sequence:

\medskip
\noindent
$\langle s(p,a),s(p',a')\rangle \equiv$ 
\begin{equation}\label{Conserve}
\equiv \lim_{n\to\infty}(\frac{1}{n}\sum_{i=1}^n s(p_i,a, t_i)s(p'_i,a',t'_i))=-u(a)\cdot u(a')\,.
\end{equation}
We refer to (\ref{Conserve}) as the \emph{Twisted Malus Law} (or \emph{TML}), where ``Twisted" correspond to the minus sign that is the trace of the structure of (\ref{Singlet}).  In particular when $a=a'$,
\begin{equation}\label{ConserveZ}
\langle s(p,a),s(p',a)\rangle\equiv -1\,,
\end{equation}
The identity (\ref{ConserveZ}) can be expressed as a \emph{conservation law}: {\emph{For the state (\ref{Singlet}) the total spin is zero along any axis $a$.}} 
%

\medskip
After Fine pointed out and used the fact that Bell's inequalities result from the fact that the extension of QM  that are needed for Bell's Theory transport us into the realm of classical statistics \cite{Fine1982a}, \cite{Fine1982b}, it was noted by Pitowsky (see\cite{Pitowsky1989a} and references therein) that Bell's Inequalities (or \emph{BIs})  (see below) are examples of inequalities that Boole discovered about one century before (in probability language) in the context of classical statistics \cite{Boole1854}.  Boole showed that \emph{Boole's Inequalities} permits one to recognize (or at lest suspect), when an inequality is not satisfied,  that sets of data supposedly indexed as the samples that these data should come from, indeed do not always come from unique samples as they should (what we loosely call \emph{Boole's Rule of Single Sample} or \emph{BRSS}). Clearly, a brutal use of QM would tell us that BIs, as special examples of Boole's inequalities that use at least three axes or four per EPRB pair cannot make physical sense: often the negative opinions expressed about Bell's Theory (see, \textit{e.g.,} \cite{HessEtal} and references therein) overlook the fact that too direct an use of Boole's theory (and in particular BRSS) in contexts where one tests aspects of \emph{``QM augmented by some form of realism"} rather than just \emph{``QM by itself"} is inappropriate. 

We notice that the perfect anti-correlation represented by (\ref{ConserveZ}) holds true for time $t$ in $[t_{00}, T_0)$ where: 

- $t_{00}$ is any time close enough to when the EPRB pair is created for no interaction to have occurred involving $p$ or $p'$ after the pair is created and up to $t_{00}$, 

\noindent
We also use: 

- $T_0=\min (t_0,t'_0)$, the first time (in the laboratory frame) when one of $p$ and $p'$ is subjected to an interaction. 

\noindent
For some $T_{00}=\max(t_0,\,t'_0)<\infty$:
 
 - $t_0\leq T_{00}$ is the first $t>t_{00}$ when $p$ is subjected to an interaction that may affect (\ref{Singlet}).

 - $t'_0\leq T_{00}$ is  the first $t>t_{00}$ when $p'$ is subjected to an interaction that may affect (\ref{Singlet}).

\emph{We are always under the \emph{``(Both Particles measured in Uniformly Bounded Time) Standing Assumption"} according to which there is some $T_{M}<\infty$ such that for all $i$'s,  $T_{00}\leq T_{M}$; then except otherwise stated, the first measurement of $p$ is along $a$ and the first measurement of $p$ is along $a'$, \textbf{by definition.}}
  
\medskip
The property ``$t$ in $[t_{00}, T_0)$" expressed above
is thus \emph{sufficient} for the symmetry (\ref{ConserveZ}), and more generally for (\ref{Conserve}) to hold true: sufficiency is all we need for our purpose. 

\smallskip
Here are 3 simplifying hypotheses:

\smallskip
\noindent
S1)  The only  \emph{possible interactions} that we consider are \emph{measurements}, except for the processes (before $t_{00}$) that is used to generate the EPRB pairs. 

\smallskip
\noindent
S2) The only  \emph{measurements} that we consider are \emph{measurements of the spin projection for the spin of spin-$\frac{1}{2}$ particles}, whether we consider these particles by themselves or as part of an EPRB pair in the Singlet state. 

\smallskip
\noindent
S3) Pairs subjected to any accident, including incomplete or otherwise spoiled measurements are ignored and in particular not indexed.

\smallskip
Assume that at times $\tau_ 0<\tau_ 1<\tau_ 2$ a triplet of measurements are made respectively along $a_j$, for each $\tau_ j$, $j$ in $\{0,1,2\}$ on particle $p$.  In any instance $i$ of such a triplet in a sequence of them, one measures three numbers $s(p,a_j, \tau_ j)$ for $j$ in $\{0,1,2\}$, and the symbol $p$ can as well be omitted as we consider a single particle only for each instance of the experiment. The first measurement \emph{prepares} $p$ (or \emph{is a preparation} of $p$) \emph{in the spin state $s(a_0, \tau_ 0)$}. This preparation statement, as we learn from QM, translates in the fact that, according to the \emph{Malus Law for spin $\frac{1}{2}$ particles}, we have $\langle s(a_0,\tau_ 0),s(a_1,\tau_ 1)\rangle=u(a_0)\cdot u(a_1)$. In particular, $\langle s(a_0, \tau_ 0),s(a_1,\tau_ 1)\rangle=1$ whenever $a_1=a_0$. For any axis $a$ we say that $p$ is \emph{$a$-prepared}  (or \emph{$a$-p} for short) at time $\tau_ 0$ if and only if, $\langle s(a,\tau_ 0),s(a',\tau_ 1)\rangle=u(a)\cdot u(a')$ whenever the measurement at $\tau_ 1>\tau_ 0$ along $a'$ is the first interaction to which $p$ is subjected after $\tau_ 0$. In particular, in the case of a Singlet state under our Standing Assumption: 

 $\quad$- The particle $p$ is $-a'$-p for $t$ in $(t_{00}, t_0)$.

 $\quad$-  The particle $p'$ is $-a$-p for $t$ in $(t_{00}, t'_0)$.

The measurement at $\tau_ 1$ can also be considered as a preparation of $p$ in the spin state $s(a_1, \tau_ 1)$, which translates \textit{mutatis mutandis} into  $\langle s(\tau_ 1), s(\tau_ 2) \rangle=u(a_1)\cdot u(a_2)$. What follows is well known in QM and experimentally:

\smallskip
\noindent 
\textbf{Erasure Lemma.} \emph{Any $a$-p spin state that could be prepared for $p$ at time $\tau_ 0$ is washed out at time $\tau_ 1$ when the measurement along $b\not = a$ is performed on $p$.}

\smallskip
We need forms of augmentation of QM that cover from the \emph{``Predictive Hidden Variables compatibles with the statistical properties of Quantum Mechanics"} (or \emph{PHVs}) used by Bell in \cite{Bell} to some weak form of ``realism" discussed \textit{e.g.,}  in \cite{Leggett2008},and in \cite{Stapp1971}, \cite{Stapp1985}, \cite{Tresser001}, and briefly below.

\smallskip
\noindent  
\textbf{Time-Realism.} We want to take into account the fact that often when one speaks colloquially  about realism in microphysics, one refers to the pre-existence of observable's values before, and in fact independently of measurement.  In general, one would then say that, according to (what we call ) \emph{Time-Realism}, any observable value (possibly up to small change) pre-exist to its own measurement.  Because the possible values of the spin projections are discrete, the Time-realist assumptions enables new meaning (using the Erasure Lemma) for the spin along $b$ at times near $t_0$ when one measures along $a$  with $t_{00}<t_0<t_{11}$ and under conditions as follows: 

\smallskip
- \emph{TR1.} In the interval $[t_0,t_{11})$ with no interaction affecting the particle during  $(t_0,t_{11})$, a value denoted by $s(p,b;a, t_0+)\equiv s(p,b, t_{11})$ accessible to all the forms of Instantaneous Realism (see below) that we consider. 

\smallskip
- \emph{TR2.} In the interval $(t_{00},t_0)$ with no interaction affecting the particle during $(t_{00},t_0)$, a value denoted by $s(p,b;a, t_0-)$ accessible only to forms of Instantaneous Realism that are predictive. 

\medskip
\noindent 
\textbf{Instantaneous Realism.}  
By \emph{Instantaneous Realism} we mean \emph{``the effect of assuming Realism on the set of observables that comprises the fact that said observables make sense and have a value, even if no experiment is made about them, even if at the same time any other observable if being measured (be it a observable that commutes or not with the one under consideration"}.  

\smallskip
- \textbf{MCD.}
As a precise form of Instantaneous Realism we will mostly use \emph{Macroscopic Counterfactual Definiteness} (or \emph{MCD}) \cite{Stapp1985}, \cite{Leggett2008}, \cite{Tresser001}, \cite{Tresser002} (see also \cite{Stapp1971} for an early version without the name): \emph{According to \emph{MCD} , if a measurement is made at time $t_0$ on $p$ along $a$ to give $s(p,a;a,t_0)$, then any measurement that one could have made instead, say along $b$ also has has a value $s(p,b;b,t_0)$ jointly with the measurement that is being made.}  

At the other end of the spectrum of realists augmentations of QM, one finds PHVs: one form of such theories is discussed with great detail in \cite{Bell}, and we will not repeat here this classical exposition which is too precise for us.  What is important for us is the following aspect of PHVs:

\smallskip
- \textbf{PHVs.}
\emph{According to \emph{PHVs}, either predictions cover exactly whatever is given meaning by  MCD, or one can use, as exotic meanings for the spin projection along $b$ at $t_0$ the two values $s(p,b;a,t_0\pm)$ of TR1 and TR2.}

\smallskip
\noindent 
\textbf{ $a$-preparation.}
Once QM is augmented, one can raise the question of whether particle $p$ can be $a$-p and $b$-p at the same time, \emph{i.e.,} \emph{``Can we find states that are both $a$-p and $b$-p for $b\not=a$? "}. The answer is given by the following easy lemma which can be proven as a trivial exercise in Boole Theory, using (\ref{V_3}) but with no room for Locality:  

\smallskip
\noindent 
\textbf{No $p$ both $a$-p and $b$-p Lemma ({\rm or} No$p$L).}{ \emph{ Let $p$ be a spin-$\frac{1}{2}$ particle. Then $p$ cannot be both $a$-prepapred and $b$-prepared at the same time for $a\not=b$.}

\smallskip
We now recall the abstract aspects of the BIs to the extent we need, covering the inequality for the case of three axes (\emph{version $V3$}) as in Bell's original paper and the inequality for four axes (\emph{version $V4$}) (also known as \emph{CHSH})  that if better suited for experimental verification.  As long as we are in an abstract setting, 3 and 4 axes for BIs mean that respectively 3 and 4 sequences of numbers in $\{-1,\, 1\}$ are considered.  We do not consider convergence issues that are discussed \textit{e.g.,} in \cite{Tresser001}.  The inequalities (\ref{V_3}) and  (\ref{V_4}), assuming convergence but still without any physical content we call \emph{Abstract Bell Inequalities} or \emph{ABIs}.  \emph{Our main objective is to show that one cannot go from ABIs to the usual BIs with their intended physical meaning (in an augmented version of QM).} 

\smallskip 
\noindent 
\textbf{ABIs.} Consider the space of quadruplets  of sequences $(\{x_i\}, \,\{y_i\}, \,\{z_i\},\,\{w_i\})$ of numbers in the set $\{-1,\,1\}$ so that the pairwise correlations exists with probability one, meaning that for $u$ and $v$ in $\{x,\,y,\,z,\, w \}$ the limits that define the correlations exist.  Then  we have:
\begin{equation}\label{V_4}
 |\langle x,y\rangle +\langle x,z\rangle| + |\langle w,y\rangle -\langle w,z\rangle|  \leq 2\,.
\end{equation}
which we call the $V4$ version of the ABIs (for short $V4$). One short proof is a tiny exercise that starts by noticing that the extremal values need to be realized when the correlation are extremal, which happens when these correlations are either $-1$ or $1$. A specialization of (\ref{V_4}) yields:
\begin{equation}\label{V_3}
| \langle x,y\rangle -\langle x,z\rangle|+\langle y,z\rangle\leq 1\,,
\end{equation}
which we call the $V3$ version of the ABIs (for short $V3$).   

\smallskip
Bell recognized in \cite{Bell} that one needs another assumption (that he qualifies indeed of crucial in \cite{Bell}), \textit{e.g.,} to compute$\langle s(p,a;a,t_0),s(p,b;b,t_0)\rangle$, Locality that can be defined as follows for \emph{QM $\wedge$ realism} (see also \cite{Shimony} for a discussion of Locality more complete than needed here). 
 
\smallskip
\noindent
\textbf {Locality.}  \emph{Locality tells us that if $(x_0, t_0)$ and $(x_1, t_1)$ are spatially separated, \textit{i.e.,}  $\Delta x ^2> c^2 \Delta t^2$, then the setting of an instrument at  $(x_0, t_0)$ cannot change the output of a measurement made at  $(x_1, t_1)$ nor the value of an observable that could be measured there and then instead of the observable that is actually being measured.}
 
\noindent
For the sake of completeness we recall a definition that we will use here only for comments on comparative statuses.
 
\smallskip
\noindent
\textbf {Effect After Cause Principle (\emph{EACP}):} \emph{For any Lorentz observer the value of an observable cannot change even  from any cause that happens after said observable has been measured for that observer.}

\smallskip
We consider first (\ref{V_3}).  Clearly, the sequences $x$, $y$ and $z$ can be paired to $s(p,a;a,t_0)$,  $s(p,b;b,t_0)$ and  $s(p,a';a',t'_0)$ in any way since the ABI works (with probability one) for any 3 sequences such that the series that define the correlations as in (\ref{Conserve}) converge. An important feature of MCD and of the most usual PHVs is that, with $a$ the axis effectively used at time $t_0$ for $p$, the form of spin projection along $b$ that is relevant is $s(p,b;b,t_0)$ and not $s(p,b;a,t_0\pm)$ that appears in an exotic interpretation of PHVs.  We take the $u\not =v$ among $x$, $y$ and $z$ so that with the pairing that is chosen, $u$ corresponds to $s(p,a;a,t_0)$ while $v$ corresponds to $s(p,b;b,t_0)$ and a third symbol $u'$ that represents then the third symbol out of $\{x,\,y,\,z\}$ and thus corresponds to $s(p',a';a',t_0)$. While the correlations $\langle u,u'\rangle$ and $\langle v,u'\rangle$ are (or so it seems but see the analysis and the Principle enunciated below) both given by QM (in both cases, we have the TML (\ref{Conserve})), what is not directly covered by QM is $\langle u,u'\rangle$ for the computation of which Bell called upon Locality. Briefly then, the usual argument is that, using Locality, $u$ does not depend upon the sequence for $p'$, so that one can write:

\smallskip
\noindent
$\langle u,v\rangle\equiv \langle s(p,a;a,t_0),s(p,b;b,t_0)\rangle=$ 
\begin{equation}\label{BellTrick1}
 =\langle - s(p,a;a,t_0),s(p',-b:-b,t'_0)\rangle=
\end{equation}
\medskip
\noindent
$\qquad\qquad\qquad\qquad=-u(a)\cdot u(-b)=u(a)\cdot u(b)\,.$ 

\noindent
using a direct extension of QM's laws to quantities that can only make sense if one uses MCD or the usual PHVs.

\noindent
\emph{However this usual reasoning is wanting}. The problem is that  $\langle u,u'\rangle$ means in effect that the particle $p$ is $-a'$-p.  But then similarly $\langle s(p,a;a,t_0),s(p',-b:-b,t'_0)\rangle=u(a)\cdot u(b)$ means in effect that the particle $p$ is also $b$-p. It then follows from the \emph{No$p$L} that the sequence $u$ as used for pairing with $v$ must be different from the sequence $u$ that is used for pairing with $u'$. We thus have a violation of \emph{BRSS} as it must be adapted to be usable in a environment where on assumes some form of realism. We have what we call a \textbf{No physics in $V3$ Lemma} that we do not formulate for now, waiting for a generalization to $V3$ and $V4$.

\smallskip  
\noindent
\textbf{Remark.} \emph{The EACP-based Bell's Theory in \cite{Tresser001},  \cite{Tresser002} uses a right angle ($u(a)\cdot u(b)=0$) and a symmetry argument to let us compute  $\langle y,z\rangle$. Thus this approach that only deals with $V3$ stands unaffected what we do here.}

But let us come back to what seemed to be the safe part of the physical interpretation of $V3$, the correlations involving $p'$. Since $p'$ is only measured along $a'$, it would seem that the \emph{No$p$L} is mute about the correlations involving $p'$. But this is only true as long as one doe not pay attention to the fact that these correlations can \emph{also} be computed using the fact that $p'$ is both: 

\smallskip
\noindent
- $-a$-p (to yield another  $\langle s(p,a;a,t_0),s(p',a';a',t'_0)\rangle$). 

\smallskip
and 

\smallskip
 \noindent
- $-b$-p (to yield another  $\langle s(p,b;b,t_0),s(p',a';a',t'_0)\rangle$).

\medskip
\noindent
Hence the \emph{No$p$L} indeed applies even in what seemed to be the easy case: the $p'$ used to compute  $\langle s(p,a;a,t_0), s(p',a';a',t'_0) \rangle$ cannot be the $p'$ used to compute $\langle s(p,b;b,t_0),s(p',a';a',t'_0)\rangle$.

\medskip
It is useful and very important not only for $V3$ but also for $V4$, that putting MCD and Locality together cannot be a nice wedding of hypotheses.  As soon as realism is effective, \textit{i.e.,} QM cannot handle by itself all the variables that are present, \emph{MCD $\wedge$ Locality} generates a configuration that leads to a contradiction because of the  \emph{No$p$L}.  For ease of use we record this status in the form of the following Principle.

\medskip
\noindent
\textbf{MCD together with Locality do not form a happy family Principle.}  \emph{In the context of EPRB pairs, if MCD is used effectively in the sense that one particle at least is endowed with more observable than QM can handle, then the combination of MCD and Locality forces one into the contradictions of the \emph{No $p$ is $a$-prepared and $b$-prepared at once Lemma} so that one particle at least must appear with two different names, invalidating the usability of the Bell Inequalities that comprise that particle.  All that applies as well when MCD is replaced by ``PHVs used as most often done".} 

\medskip
\noindent
By a direct application of this Principle, we can see that $V4$ does not produce more contradictions when using Locality that when assuming Non-locality.  But the importance of this Principle does not lie in the fact that we do not need to repeat the argument used for $V3$ in order to compute, \textit{e.g.,} the correlation in $V4$ between the two axes that can only be treated by assuming some form of realism. What is important is the following state of affairs: if that special pair of axes would be the only problem, as a quick analysis seems to suggest that it was (essentially) the case when first considering $V3$, we would have three terms that would be accessible to fair and unique evaluation in  (\ref{V_4}).  Assume then that, \textit{e.g.,} some symmetry argument would let us have good enough bounds on the supposedly ``only problematic link" of the chain of correlations. Then one would still be able to use  (\ref{V_4}) to test the conjunction MCD$\,\wedge \,$Locality so that altogether \emph{we would only have here a quantitative modification of Bell's Theory.} The above Principle, tell us to the contrary that out of  (\ref{V_4}), only the correlation that makes sense in ordinary QM can be computed. The conclusion is thus very dramatic: \emph{Boole-Bell Theory cannot help us study the conjunction MCD$\,\wedge\,$Locality as is usually done.} 

\smallskip
\noindent
\textbf{No physics in $V3$ nor $V4$ Lemma.}
\emph{One cannot give a physical meaning to the $V3$ nor $V4$ versions of the ABIs on the basis of adding Locality to MQ and to the MCD hypotheses. Indeed, only the correlation between the spin projections that are actually measured, hence compatible with ordinary QM,  can be uniquely evaluated by reduction to known physical laws.} 

\smallskip
The only remaining hope for preserving Bell's Theory based on BI's lies in exotic utilization of PHVs.   However a careful analysis reveals that the significations for $s(p,b;a,t_0\pm)$ from TR1 and TR2 (that are seldom used in Bell's Theory papers anyway) do not help us to give a meaning to $V3$ nor $V4$. In studying the exotic valuations, we uses in particular again the \emph{No$p$L}: details and more on that and other matters (including more forms of realism, Bell's Theory without inequalities, and references to works of Leggett, Scully, etc. parallel to -but different from- ours) will be provided elsewhere.


\begin{thebibliography}{99}

\bibitem{EPR}
A.\ Einstein, B.\ Podolsky, N.\ Rosen \textit{Phys. Rev.} \textbf{47}, 777 (1935).

\bibitem{Bohm}
D.\ Bohm \textit{Quantum Theory,} (Prentice Hall; New York 1951).

\bibitem{Bell}
J.S.\ Bell, \textit{Physic} (Long Island City, NY)  \textbf{1}, 195 (1964).

\bibitem{CHSH69}
J.F.\ Clauser, M.A.\ Horne, A.\ Shimony, R.A.\ Holt, \textit{Phys.
Rev. Lett.} \textbf{23}, 880 (1969).

\bibitem{Wigner70} 
E.P.\ Wigner, \textit{ Am. J. Phys.} \textbf{38}, 1005 (1970). 

\bibitem{Stapp1971} 
H.P.\ Stapp, \textit{Phys. Rev.} \textbf{3 D}, 1303 (1971).

\bibitem{Shimony}
A.\ Shimony, in \textit{The Stanford Encyclopedia of Philosophy,}  E.N. Zalta, Ed.  (2009), {\url{http://plato.stanford.edu/archives/sum2009/entries/bell-theorem/}}. 

\bibitem{Le Bellac}
M. Le Bellac \textit{Quantum Physics,}  (Cambridge University Press; Cambridge 2006).

\bibitem{AspectEtAl1982}
A.\ Aspect, J.\ Dalibard, G.\ Roger, \textit{Phys. Rev. Lett.} \textbf{49}, 1804 (1982).

\bibitem{Fine1982a}
A.\ Fine, \textit{Phys. Rev. Lett.} \textbf{48}, 291 (1982).

\bibitem{Fine1982b}
A.\ Fine, \textit{J. Math. Phys.} \textbf{23}, 1306 (1982).

\bibitem{Pitowsky1989a}
I.\ Pitowsky, in \textit{Bell Theorem, Quantum Theory and the Conception of the Universe,} M. Kafatos, Ed.  (Kluwer; Dordrecht, 1989). 

\bibitem{Boole1854}
G.\ Boole, \textit{Phil. Trans. Roy. Soc. (London)}  \textbf{152}, 225 (1862).

\bibitem{HessEtal}
K. \ Hess, K.\ Michielsen2 , H.\ De Raedt \textit{EPL}, \textbf{87},  60007, (2009).

\bibitem{Tresser001}
C.\ Tresser, \textit{Eur. Phys. J. D.}  \textbf{58}, 385 (2010).

\bibitem{Tresser002}
C.\ Tresser, \textit{Bell's Theorem with no Locality asumption: putting Free Will at work}, Preprint (2010). 

\bibitem{Leggett2008}
A.J.\ Leggett, \textit{Rep. Prog. Phys.} \textbf{71}, 022001 (2008).

\bibitem{Stapp1985}
H.P.\ Stapp, in \textit{Symposium on the Foundations of Modern Physics: 50 years of the Einstein-Podolsky-Rosen Gedankenexperiment,}  P. Lahti and P. Mittelstaedt, Eds.  (World Scientific, Singapore, 1985), pp. 637Ð-652. 

\end{thebibliography}
\end{document}